\documentclass[preprint,12pt]{elsarticle}

\usepackage{amsmath,amssymb}
\usepackage[english]{babel}
\usepackage{natbib}

\newcommand{\pref}[1]{(\ref{#1})}

\newcommand{\Eq}[1]{Eq.~(\ref{#1})}

\DeclareMathOperator*{\eins}{\mathbf{1}}
\DeclareMathOperator*{\Mor}{\mathrm{Mor}}
\DeclareMathOperator*{\0}{\mathbf{0}}

\bibpunct{(}{)}{;}{}{,}{;}

\journal{Topics in Cognitive Science}

\begin{document}

\bibliographystyle{apalike}


\begin{frontmatter}

\title{\bf Order effects in dynamic semantics}

\author{Peter beim Graben\corref{cor1}}
\address{Department of German Language and Linguistics \\
 Humboldt-Universit\"at zu Berlin,
 Germany}
\ead{peter.beim.graben@hu-berlin.de}
\ead[url]{www.beimgraben.info}
\cortext[cor1]{Department of German Language and Linguistics \\
 Humboldt-Universit\"at zu Berlin \\
 Unter den Linden 6 \\
 D -- 10099 Berlin \\
 Phone: +49-30-2093-9632 \\
 Fax: +49-30-2093-9729
 }

\begin{abstract}
In their target article, \citet{WangBusemeyer13} [A quantum question order model supported by empirical tests of an a priori and precise prediction. \emph{Topics in Cognitive Science}] discuss question order effects in terms of incompatible projectors on a Hilbert space. In a similar vein, Blutner recently presented an orthoalgebraic query language essentially relying on dynamic update semantics. Here, I shall comment on some interesting analogies between the different variants of dynamic semantics and generalized quantum theory to illustrate other kinds of order effects in human cognition, such as belief revision, the resolution of anaphors, and default reasoning that result from the crucial non-commutativity of mental operations upon the belief state of a cognitive agent.
\end{abstract}

\begin{keyword}
Question order effects, belief revisions, anaphor resolution, default reasoning, generalized quantum theory, dynamic semantics
\end{keyword}

\end{frontmatter}

\section{Introduction}
\label{sec:intro}

\noindent
In their target article, \citet{WangBusemeyer13} discuss question order effects in terms of incompatible, i.e. non-commuting, projectors on a Hilbert space. In their model, a person's belief state is expressed by a vector in a linear space that is equipped with a scalar product while the answers to yes/no-questions correspond to orthogonal subspaces in Hilbert space. A question is answered by projecting the current belief state vector either onto the question's \emph{yes} or \emph{no} subspace. When answer subspaces to different questions do not coincide, the sequence of projections matters and questions are incompatible to each other.

In a similar vein, \citet{Blutner12} presented an (ortho-) algebraic approach for a query language that is not only able to explain question order effects but also allows the analysis of conditional questions of the form ``If Mary reads this book, will she recommend it to Peter?'' \citep{Blutner12}. Also Blutner's approach essentially relies upon a Hilbert space representation of belief states where questions induce a decorated partition into orthogonal answer subspaces. Yet, \citet{Blutner12} explicitly constructs his query language as a ``version of update semantics'' \citep{Blutner96, Veltman96} where the ``meaning of a sentence is not its truth condition but rather its impact on the hearer'' \citep{Kracht02}.

In my commentary on the target article of Wang and Busemeyer, I shall further elaborate the interesting analogies between the different variants of \emph{update semantics} \citep{Blutner96, Veltman96}, \emph{dynamic semantics} \citep{Gaerdenfors88, Graben06, Kracht02}, and \emph{dynamic logics} \citep{GroenendijkStokhof91, Staudacher87} on the one hand and \emph{generalized quantum theory} \citep{AtmanspacherRomerWalach02}, respective \emph{quantum dynamic logic} \citep{BaltagSmets11} on the other hand in order to illustrate some other kinds of order effects in human cognition, such as belief revision, the resolution of anaphors, and default reasoning that essentially result from the non-commutativity of mental operations upon a person's belief states.

\section{Generalized Quantum Theory}
\label{sec:gqt}

\noindent
In generalized (or ``weak'') quantum theory, \citet{AtmanspacherRomerWalach02} consider a set $X$ as a general state space and functions (\emph{morphisms}) $\Mor(X) = \{ A | A : X \to X \}$, transforming a state $x \in X$ into another state $y \in X$ through
\begin{equation}\label{map}
    y = A(x) \:.
\end{equation}
Particular functions from a subset $\mathcal{A} \subseteq \Mor(X)$ are called \emph{observables}. Observables can be concatenated, i.e. iteratively invoked, such that $(B \circ A)(x) = B(A(x)) = B(y)$, for all $x \in X$. This \emph{observable product} $AB = A \circ B$ is associative: $A(BC) = (AB)C$, but in general not commutative: $AB  \ne BA$. Only when $AB = BA$, observables are called \emph{compatible}, otherwise they are called \emph{incompatible}.

\Citet{AtmanspacherRomerWalach02} supply a number of further axioms describing the properties of such observables and their impact upon the state space $X$. One of these axioms introduces a neutral element $\eins$, such that
\begin{equation}\label{tauto}
    \eins \circ A = A \circ \eins = A
\end{equation}
for all $A \in \mathcal{A}$. Another axiom additionally introduces a zero observable $\0 \in \mathcal{A}$ and a zero state $o \in X$, such that
\begin{eqnarray}
  \0(x) &=& o \\
  A(o) &=& o \\
  \0 A &=& A \0 = \0
\end{eqnarray}
for all $x \in X$ and $A \in \mathcal{A}$.

An important class of observables $\mathcal{P} \subset \mathcal{A}$ are \emph{projectors} which are \emph{idempotent}
\begin{equation} \label{idempotent}
    A^2 = A A = A \:.
\end{equation}
Applying a projector $A$ to a state $x \in X$ yields another state $y = A(x) = A^2(x) = A(A(x)) = A(y)$ that does not change under subsequent applications of $A$ anymore. The projected state $y = A(y)$ is hence an eigenstate of $A$.

\section{Classical Dynamic Semantics}
\label{sec:dyse}

\noindent
Regarding the state space $X$ of generalized quantum theory as the set of epistemic states of a cognitive agent, yields an instantiation of \emph{dynamic update semantics} \citep{Blutner96, Gaerdenfors88, Veltman96} in the following way: Elements $x, y, z \in X$ are called \emph{epistemic states}, or \emph{belief states} while observables $A, B \in \mathcal{A}$ become interpreted as \emph{epistemic operators}. By restricting observables only to commutative and idempotent operators, one obtains \emph{propositions}. Their (commutative) composition can then be identified with logical \emph{conjunction}
\begin{equation}\label{Gaerdenfors:conj}
    A \wedge B = AB = BA = B \wedge A \:.
\end{equation}

An important notion in dynamic semantics is that of \emph{acceptance}. A proposition $A \in \mathcal{P}$ is said to be \emph{accepted in state} $x \in X$ (or $x$ is accepting $A$), if
\begin{equation}\label{Gaerdenfors:accept}
    A(x) = x \:.
\end{equation}
That means, the state $x$ is an eigenstate of $A$. Because propositions are idempotent, the state $y = A(x)$ always accepts $A$. Thus, \Eq{map} receives a straightforward interpretation as information update.

Furthermore, \emph{logical consequence} (or stability in \citet{Blutner96}) is defined as follows: A proposition $B$ is called a logical consequence of a proposition $A$, if
\begin{equation}\label{Gaerdenfors:consqeq}
    B \wedge A = A \wedge B = A \:.
\end{equation}
In this case, $y = A(x)$ entails $B(y) = B(A(x)) = A(x) = y$, such that $B$ is accepted whenever $A$ is accepted in an epistemic state (but not vice versa).

The given system can be equipped with other logical connectives such as negation ($\neg A$) or disjunction
($A \vee B$). \Citet{Gaerdenfors88} has proven that the resulting calculus is equivalent to intuitionist logics which can be further extended to classical propositional logics. Another important extension is Bayesian update semantics where states are interpreted as probability distributions $\rho$ over propositions. Then the impact of a proposition $A$ upon a belief state $\rho$ is expressed by Bayesian conditionalization
\begin{equation}\label{conditionalize}
    \rho_A(B) = \frac{\rho(B \wedge A)}{\rho(A)} =: \rho(B | A)
\end{equation}
of the distribution with respect to $A$ \citep{BenthemGerbrandyKooi09, Gaerdenfors88, Graben06}.

\section{Non-classical Dynamic Semantics}
\label{sec:ncdyse}

\noindent
Classical dynamic semantics comprises epistemic operators that are commutative and idempotent propositions. Moreover, such systems are monotonic, as propositions which already have been accepted remain accepted during the updating of epistemic states. This follows from commutativity: Let $A$ be  accepted in state $x$ (i.e. $A(x) = x$) and let $B(x) = y$, such that $B$ is learned during the updating from $x$ to $y$. Then $A(y) = A(B(x)) = (A \wedge B)(x) = (B \wedge A)(x) = B(A(x)) = B(x) = y$, saying that $A$ and $B$ are both accepted in the updated state $y$.

\subsection{Belief revision}
\label{sec:revis}

\noindent
However, this account is not appropriate when belief states have to be revised by new evidence. Belief revision processes are in general not commutative and hence non-monotonic such that order effects become ubiquitous. \Citet{Gaerdenfors88} introduces a belief-revision operator as a mapping $* : \mathcal{P} \to \mathcal{A} \setminus \mathcal{P}$ assigning a revision $A^* \in \mathcal{A} \setminus \mathcal{P}$ to a proposition $A \in \mathcal{P}$. This revision dynamics has to obey several minimality axioms.

In order to illustrate this process, consider an agent in a belief state $x$ that accepts the proposition $A=$``the moon consists of blue cheese'' \citep{Graben06}. Its revision is hence $A^*=$ ``the moon does not consist of blue cheese''. Another proposition might be $B=$ ``the moon consists of stone''. Since $x$ accepts $A$, the application of $B$, $B(x)$, leads to the zero state $o \in X$ of generalized quantum theory, that becomes now interpreted as the \emph{absurd state} accepting every proposition \citep{Gaerdenfors88}. Therefore also $BA=$ ``the moon consists of blue cheese and of stone'' is accepted in $o$. This state does not change under the revision $A^*$, hence $(A^* B)(x) = o$. On the other hand the product $BA^*$ applied to $x$ yields $B(y)$ where $y = A^*(x)$ accepts the revision of $A$. Therefore, $BA^*(x) \ne o$ because $BA^*=$ ``the moon does not consist of blue cheese, it rather consists of stone'' can be consistently accepted. Thus $A^*B \ne BA^*$, i.e.~belief revisions and propositions do generally not commute and are hence incompatible to each other.

Belief revision in a probabilistic, Bayesian framework requires conditionalization with respect to a minimally altered probability distribution $\rho^*$ which involves several technical peculiarities such as epistemic entrenchment \citep{BaltagSmets08, Gaerdenfors88, Graben06}. In the framework of quantum cognition, however, Bayesian conditionalization is replaced by the L\"uders-Niestegge rule \citep{Luders51, Niestegge08}
\begin{equation}\label{niestegge}
    \rho_A(B) = \frac{\rho(A B A)}{\rho(A)}
\end{equation}
(see also \citet{Blutner09, BlutnerPothosBruza13}) resulting from the non-commutativity of Hilbert space projections \citep{AtmanspacherRomerWalach02}, where $\rho(A) = |\langle \psi | A | \psi \rangle|^2$ gives the quantum probability in state vector $| \psi \rangle$. Therefore, belief revision seems to be a good candidate for quantum probability models in dynamic semantics \citep{EngesserGabbay02}.\footnote{
    Equation \pref{niestegge} holds for self-adjoint projectors on Hilbert space. For general operators that are not necessarily self-adjoint, one had \citep{AtmanspacherRomerWalach02}
    \[
            \rho_A(B) = \frac{\rho(A^* B A)}{\rho(A^* A)} \:.
    \]
    It might be tempting to speculate about the possible relationship between the belief revision operator ``$*$'' in dynamic semantics and the algebra involution ``$*$'' in quantum theory.
}

\subsection{Anaphor resultion}
\label{sec:ana}

\noindent
Another important example for order effects in dynamic semantics is the resolution of anaphors. For this aim, \citet{Staudacher87} and \citet{GroenendijkStokhof91} have independently developed models of dynamic predicate logics, where quantifiers, such as ``there exists an $x$'' or ``for all $x$'', and anaphors, e.g. pronouns, are described by epistemic operators acting upon model theoretic valuations (see also \citet{Kracht02}).

As an instructive example we consider three propositions $A=$``John sat at the table'', $B=$``George came in'', and $C=$``he was wearing a hat'' \citep{Graben06}. Here, the pronoun ``he'' assumes conflicting interpretations for the compositions $CBA=$ ``John sat at the table; George came in; he was wearing a hat'' and $CAB$= ``George came in; John sat at the table; he was wearing a hat''. In the first case, the pronoun ``he'' refers to ``George'', while it refers to ``John'' in the second case. These anaphors therefore have to be described as non-commutative operators as well.

\subsection{Default reasoning}
\label{sec:defaults}

\noindent
Finally, \citet{Blutner96} and \citet{Veltman96} have observed that by relaxing the stability condition of logical consequence in \pref{Gaerdenfors:consqeq}, dynamic logics becomes non-monotonic. This allows the treatment of default operations, such as ``may'' or ``normally''. \Citet{Veltman96} presented a nice example for such an ordering effect in default reasoning: Let $A=$``Somebody is knocking at the door'', $B=$``Maybe it's John'', and $C=$``It's Mary''. Then the composition $CBA=$``Somebody is knocking at the door. Maybe it's John. It's Mary.'' makes perfect sense, while $BCBA=$``Somebody is knocking at the door. Maybe it's John. It's Mary. Maybe it's John.'' does not.

\section{Conclusion}
\label{sec:concl}

\noindent
Classical dynamic semantics formalizes propositional logics in terms of commutative and idempotent epistemic operators that constitute a monotonic system of belief updating dynamics. By contrast, belief revision, the resolution of anaphors and non-monotonic reasoning in default logics require non-commutative operations.

In probabilistic dynamic semantics, updating is expressed by means of Bayesian conditionalization, whereas the description of belief revision processes requires rather peculiar mechanisms that could probably be more naturally expressed by means of quantum probability theory.

Other types of cognitive order effects such as anaphor resolution or default reasoning have been successfully described by extensions of dynamic semantics including predicate calculus or non-monotonicity. To my present knowledge, probabilistic generalizations of these models utilizing quantum probability theory have not yet been devised. This might be a promising direction for future research.

\section*{Acknowledgements}

I gratefully acknowledge support from the German Research Foundation (DFG) through Heisenberg fellowship GR 3711/1-1 and by the Franklin Fetzer Trust.



\end{document}